\documentclass[letterpaper,journal]{IEEEtran}
\usepackage{graphicx}
\usepackage{latexsym}
 \usepackage{amsmath}
 \usepackage{amssymb}
 \usepackage{bm}
 \usepackage{amsmath,amssymb}
 \usepackage{graphicx}
 \newcommand{\tentative}{\noindent{{\underline{\tt tentative decision}}} }

\newcommand{\ctov}{\noindent{{\underline{\tt check to variable}}} }
\newcommand{\vtoc}{\noindent{{\underline{\tt variable to check}}} }

\newcommand{\initialization}{\noindent{{\underline{\tt initialization}}} }

\newcommand{\GF}{{\mathrm{GF}}}
\newcommand{\GL}{\mathrm{GL}}
\newcommand{\RR}{{\mathbb R}}

\newcommand{\gb}[2]{\left[ \begin{array}{c} #1\\#2\end{array}\right]}

\title{
Multiplicatively Repeated Non-Binary LDPC Codes
}

\author{
Kenta~Kasai, 
David~Declercq, 
Charly~Poulliat,~\IEEEmembership{Member,~IEEE,} 
and~Kohichi~Sakaniwa,~\IEEEmembership{Senior Member,~IEEE}
 \thanks{The material in this paper was presented in part at 2010 {IEEE} International Symposium on  Information Theory (ISIT) \cite{isit_nb_lr}. 

K.~Kasai and K.~Sakaniwa are with the Department of Communications and Integrated
Systems, Graduate School of Science and Engineering, Tokyo Institute of
Technology, 2-12-1 Ookayama, Meguro-ku, Tokyo 152-8550, Japan (e-mail: \{kenta, sakaniwa\}@comm.ss.titech.ac.jp).

D.~Declercq and C.~Poulliat are with the ETIS Laboratory, UMR8051 ENSEA UCP CNRS,
95014 Cergy-Pontoise, France (e-mail: \{declercq, poulliat\}@ensea.fr).
}
}

\begin{document}
\maketitle
\begin{abstract}
We propose non-binary LDPC codes concatenated with multiplicative repetition codes. 
By multiplicatively repeating the (2,3)-regular non-binary LDPC  mother code of rate 1/3, 
we construct rate-compatible codes of lower rates $1/6, 1/9, 1/12, \dotsc$.
Surprisingly, such simple low-rate non-binary LDPC codes outperform the best low-rate binary LDPC codes so far. 
Moreover, we propose the decoding algorithm for the proposed codes, which can be decoded  
with almost the same computational complexity as that of the mother code. 
\end{abstract}

\section{Introduction}
\label{150913_18Feb10}
In 1963, Gallager invented low-density parity-check (LDPC) codes \cite{gallager_LDPC}.
Due to sparsity of the code representation, 
LDPC codes are efficiently decoded by belief propagation (BP) decoders. 
By a powerful optimization method {\it density evolution} \cite{910577}, developed by Richardson and Urbanke,
messages of BP decoding can be statistically evaluated. 
The optimized LDPC codes can approach very close to Shannon limit \cite{richardson01design}.

Rate-adaptability is a desirable property of coding systems. 
Over time-varying channels, the system adapts the coding rate according to the quality of the channels. 
Using the different type of codes for different rates results in a complex coding system. 
It is desirable to use a single encoder and decoder pair compatible with different rates. 
Such a property of codes is referred to as rate-compatibility. 
Moreover, rate-compatible codes allow us to transmit bits gradually in conjunction with
automatic repeat request (ARQ). By puncturing a low rate code, we can construct rate-compatible codes of higher rates. 

In order to reliably transmit information over the very noisy communication channels, one needs to encode the information at low coding rate. 
As described in \cite{met}, one encounters a  difficulty when designing low-rate LDPC codes. 
While, for high rate codes, even binary regular LDPC codes have good thresholds. 
The optimized low-rate structured LDPC codes, e.g.~accumulate repeat accumulate (ARA) code \cite[Table.~1]{1523619} of rate $1/6$ and multi-edge type LDPC code \cite[Table.~X]{met} of rate $1/10$ have good thresholds. 
However, the maximum row-weights of those codes are as high as 11 and 28, respectively. 
Such high row weights lead to dense parity-check matrices and  degraded performance for short code length. 
We note that, with very large code length,  
generalized LDPC codes with Hadamard codes \cite{4106145} perform very close to  the ultimate Shannon limit \cite{haykin01}. 
However, the large code length leads to transmission latency. 
If two error correcting codes with the same error-correcting capabilities and different code length are given, the shorter code is preferred. 

Another obstacle blocking the realization of the low-rate LDPC codes
is  the large number of check node computations. 
For a fixed information length $K$, it can be easily seen that the number $M$ of check nodes  gets larger as the coding rate $R$ gets lower. To be precise, $M=K(1-R)/R$. 
In the BP decoding, computations of check nodes are usually more complex than those of variable nodes. 
It is a desirable property for the low-rate LDPC codes to be decoded with computational complexity comparable to that of the higher-rate LDPC codes.

The problems for constructing low-rate LDPC codes are summarized as follows. 
\begin{itemize}
 \item {\bfseries Problem 1}: The Tanner graphs of low-rate LDPC codes tend to have many check nodes that require more complex computations than variable nodes.
 \item {\bfseries Problem 2}: The Tanner graphs of optimized low-rate LDPC codes tend to have check nodes of high degree, which results in  the degraded decoding performance for small code length. 
 \item {\bfseries Problem 3}: The optimized low-rate LDPC codes need to be used with large code length to exploit the potential decoding performance. 
\end{itemize}
In this paper, we deal with all these issues. 

In this paper, we consider non-binary LDPC codes defined by sparse parity-check matrices over $\GF(2^m)$ for $2^m>2$. 
Non-binary LDPC codes were invented by Gallager \cite{gallager_LDPC}. 
Davey and MacKay \cite{DaveyMacKayGFq} found non-binary LDPC codes can outperform binary ones. 
Non-binary LDPC codes have captured much attention recently due to their decoding performance \cite{4717467,1512327,5089504,5191110,4939224}. 

It is known that irregularity of Tanner graphs help improve the decoding performance of binary LDPC codes \cite{richardson01design}. 
While, it is not the case for the non-binary LDPC codes. 
The $(2, k)$-regular non-binary LDPC codes over $\GF(2^m)$ are empirically known \cite{4641893} as the best performing codes for $2^m\ge 64$, especially for short code length. 
This means that, for designing non-binary LDPC codes, one does not need to optimize the degree distributions of Tanner graphs, since $(2, k)$-regular non-binary LDPC codes are best. 
Furthermore, sparsity of $(2, k)$-regular Tanner graph helps efficient decoding. 

Sassatelli et al.~proposed {\itshape hybrid non-binary LDPC codes} \cite{4658702} whose symbols are defined over the Galois fields of different sizes, e.g.~over $\GF(2), \GF(8),$ and  $\GF(16)$ and 
whose Tanner graphs are irregular. In other words, the codes have two types of irregularity, i.e. irregularity of the degree distributions of graphs and the size distributions of Galois fields. 
To the best of the authors' knowledge, the decoding performance of the hybrid non-binary LDPC codes are best so far among the low-rate codes of short code length. 


In this paper, we investigate non-binary LDPC codes concatenated with multiplicative repetition inner codes. 
We use a (2, 3)-regular non-binary LDPC code of rate 1/3, as a mother code.  
By multiplicatively repeating the mother code, 
we construct codes of lower rates 1/6,1/9,1/12,$\dotsc$.
Furthermore, we present a decoding algorithm for the proposed codes. 
And we show the computational complexity of decoding is almost the same as that of the mother code. 
The codes exhibit surprisingly better decoding performance than the best codes so far for small and moderate code length. 

The rest of this paper is organized as follows. 
Section \ref{223745_17Feb10} defines the proposed codes. 
Section \ref{223813_17Feb10} describes the decoding algorithm for the proposed codes. 
In Section \ref{224134_17Feb10}, we investigate the thresholds for the proposed codes transmitted over the 
binary erasure channels (BEC) by {\itshape density evolution} \cite{richardson01design,RaU05/LTHC}. 
In Section \ref{224143_17Feb10}, 
for the BEC and AWGN channels, 
we compare the decoding performance of the proposed codes and the best known codes  for short and moderate code length. 
\section{Concatenation of Non-Binary LDPC Codes and Multiplicative Repetition Codes}
\label{223745_17Feb10}
We deal with elements of $\GF(2^m)$ as non-binary symbols. 
For transmitting over the binary input channels, each non-binary symbol in $\GF(2^m)$ needs to be represented by a binary sequence of length $m$.
For each $m$, we fix a Galois field $\GF(2^m)$ with a primitive element $\alpha$ and its primitive polynomial $\pi$. 
Once a primitive element $\alpha$ of $\GF(2^m)$ is fixed, each symbol is given a $m$-bit representation \cite[pp.~110]{macwilliams77}.
For example, with a primitive element $\alpha\in\GF(2^3)$ such that $\pi(\alpha)=\alpha^3+\alpha+1=0$, each symbol is represented as
$0=(0,0,0)$, $1=(1,0,0)$, $\alpha=(0,1,0)$, $\alpha^2=(0,0,1)$,
$\alpha^3=(1,1,0)$, $\alpha^4=(0,1,1)$, $\alpha^5=(1,1,1)$ and $\alpha^6=(1,0,1)$.

A non-binary LDPC code $C$ over $\GF(2^m)$ is defined by the null space of a sparse $M\times N$ parity-check matrix $H=\{h_{ij}\}$ defined over $\GF(2^m)$. 
\begin{align*}
 C&=\{x\in \GF(2^m)^N\mid Hx=0\in\GF(2^m)^M\}
\end{align*}
The $c$-th parity-check equation for $c=1,\dotsc,M$ is written as
\begin{align*}
h_{c1}x_1+\cdots+h_{cN}x_N=0\in\GF(2^m),
\end{align*}
where $h_{c1},\dotsc, h_{cN}\in \GF(2^m)$ and $x_1,\dotsc, x_N\in\GF(2^m)$. 


\begin{figure}[t]
\begin{center}
  \includegraphics[scale=1.1]{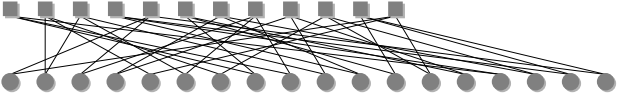}
 \vspace{-15mm}
 \caption{An example of a mother code $C_1$.  A  non-binary (2,3)-regular LDPC code of rate 1/3 over $\GF(2^m)$. 
 Each variable node represents a symbol in $\GF(2^m)$.
 Each check node represents a parity-check equation over $\GF(2^m)$.
 The code length is 18 symbols in $\GF(2^m)$ or equivalently $18m$ bits.  
Circle and square nodes represent variable and check nodes, respectively.
The lower-rate codes $C_T$ for $T=2,3\dotsc$ are constructed from $C_1$. 
 }
 \label{203721_13Jan10}
\end{center}
\end{figure}

Binary LDPC codes are represented by Tanner graphs with variable and check nodes \cite[pp.~75]{mct}. 
The non-binary LDPC codes, in this paper, are also represented by bipartite graphs with variable nodes and check nodes, which are also referred to as Tanner graphs. 
For a given sparse parity-check matrix $H=\{h_{cv}\}$ over $\GF(2^m)$,  the graph is defined as follows. 
The $v$-th variable node and $c$-th check node are connected if  $h_{cv}\neq 0$. 
By $v=1,\dotsc,N$ and $c=1,\dotsc,M$, we also denote the $v$-th variable node and $c$-th check node, respectively. 

A non-binary LDPC code with a parity-check matrix over $\GF(2^m)$ is called $(d_v, d_c)$-regular if 
all the columns and all the rows of the parity-check matrix have  weight  $d_v$ and $d_c$, respectively,  
or equivalently all the variable and check nodes have degree $d_v$ and $d_c$, respectively.

Let $C_1$ be a $(2,3)$-regular LDPC code defined over $\GF(2^m)$ of length $N$ symbols or equivalently $mN$ bits and  of rate $1/3$.
The code $C_1$ has a $2N/3\times N$  sparse parity-check matrix $H$ over $\GF(2^m)$. The matrix $H$ has row weight 3 and column weight 2. 
Fig.~\ref{203721_13Jan10} shows the Tanner graph of an example $C_1$ of length $N=$18 symbols. 

\begin{figure}[t]
\begin{center}
  \includegraphics[scale=1.1]{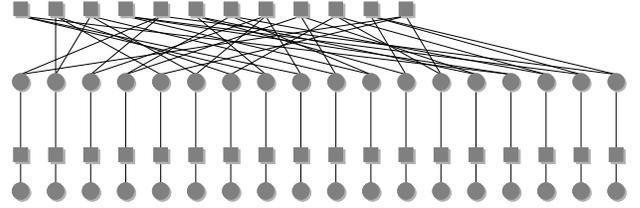}
 \caption{An example of $C_2$. A non-binary (2,3)-regular LDPC code over $\GF(2^m)$ concatenated with inner multiplicative repetition codes of length 2. 
 The code length is 36 symbols or equivalently $36m$ bits.
 The rate is 1/6. }
 \label{174554_8Jan10}
\end{center}
\end{figure}

By using  $C_1$ as a mother code, we will construct codes $C_2, C_3,\dotsc, C_T$ of lower rates in the following way. 
Choose $N$ coefficients $r_{N+1},\dotsc, r_{2N}$ uniformly at random  from $\GF(2^m)\setminus\{0\}$.
The lower-rate code $C_2$ is constructed as follows. 
\begin{align*}
 C_2=\{ &(x_1,  \dotsc, x_{2N})|x_{N+v}=r_{N+v}x_v, \\
&\quad \text{ for } v=1,\dotsc,N, (x_1, \dotsc, x_N)\in C_1\}. 
\end{align*}
Since the resulting code $C_2$ has code length $2N$  and the same number of codewords as $C_1$, then the rate is $1/6$.
Fig.~\ref{174554_8Jan10} shows the Tanner graph of $C_2$ of length 2$N=$36 symbols. 
We say that $x_{N+v}=r_{N+v}x_v$ is a {\it multiplicative repetition} symbol of $x_v$ for $v=1,\dotsc,N$. 
Each variable node of degree one in Fig.~\ref{174554_8Jan10} represents a multiplicative repetition symbol $x_{N+v}$ for $v=1,\dotsc,N$. 
And each check node of degree two in Fig.~\ref{174554_8Jan10} represents a parity-check constraint  $x_{N+v}+r_{N+v}x_v=0$ for $v=1,\dotsc,N$. 

\begin{figure}[t]
\begin{center}
 \thicklines
 \begin{picture}(205,70)(65,25)
 \put(65,83){\scriptsize{SOURCE}}
 \put(65,78){\vector(1,0){40}}
 \put(107,73){\framebox(15,15){\scriptsize{$C_1$}}}
 \put(215,85){\scriptsize{$(x_1,\dotsc,x_N)$}}
 \put(215,65){\scriptsize{$(r_{N+1}x_1,\dotsc,r_{2N}x_N)$}}
 \put(215,45){\scriptsize{$(r_{2N+1}x_1,\dotsc,r_{3N}x_N)$}}
 \put(123,78){\vector(1,0){170}}
 \put(150,38){\line(0,1){40}}
 \put(170,54){\framebox(35,15){\scriptsize{$\substack{\!\!\!\!\!\!\!\!{r_{N+v}}\\ (v=1,\dotsc,N)}$}}}
 \put(170,34){\framebox(35,15){\scriptsize{$\substack{\!\!\!\!\!\!\!\!{r_{2N+v}}\\ (v=1,\dotsc,N)}$}}}
 \put(150,58){\vector(1,0){20}}
 \put(150,38){\vector(1,0){20}}
 \put(205,58){\vector(1,0){87}}
 \put(205,38){\vector(1,0){87}}
 \end{picture}
 \caption{
 The block diagram of the encoder of $C_3$. 
 First, source of $N/3$ symbols in $\GF(2^m)$ are encoded with a (2,3)-regular LDPC code $C_1$ over $\GF(2^m)$. 
 Next, each symbol in the codeword $x_v$, for $v=1,\dotsc,N$,   is randomly multiplied by $r_{N+v}$ and $r_{2N+v}$ from $\GF(2^m)\setminus\{0\}$ 
 to generate $x_{N+v}$ and $x_{2N+v}$. 
 }
 \label{094020_13Jan10}
\end{center}
\end{figure}
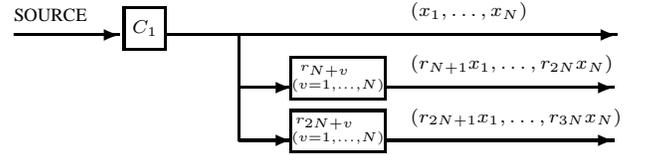

For $T\ge 3$, in a recursive fashion,  by choosing $N$ coefficients $r_{(T-1)N+1},\dotsc, r_{TN}$ randomly chosen from $\GF(2^m)\setminus\{0\}$, 
the further low-rate code $C_T$ is constructed from $C_{T-1}$ as follows. 
\begin{align*}
 C_T=\{ &(x_1,  \dotsc, x_{TN})|x_{(T-1)N+v}=r_{(T-1)N+v}x_v, \\
&\text{ for } v=1,\dotsc,N,  (x_1, \dotsc, x_{(T-1)N})\in C_{T-1}\}. 
\end{align*}
The code $C_T$ has length $TN$ and rate $1/(3T)$. 
Fig.~\ref{175738_8Jan10} shows the Tanner graph of $C_3$ of 3$N=$54 symbol code length. 
Fig.~\ref{094020_13Jan10} shows the block diagram of the encoding of $C_3.$
We refer to $T$ as {\itshape the repetition parameter}.

Concatenating a binary code with repetition codes is known as the worst coding scheme. 
Indeed, repeating a binary code just doubles the number of channel use without any improvement 
of the curve of the decoding error rate v.s. $\mathrm{E_b/N_0}$. 
Note that the proposed code $C_T$ are not generated by 
simple repetitions of the mother code but the random multiplicative repetitions of non-binary symbols. 
Since the coefficient $r_{N+v}, \dotsc, r_{(T-1)N+v},\in\GF(2^m)\setminus\{0\}$ is randomly chosen, the multiplicative repetition $x_v \mapsto (x_v, r_{N+v} x_v, \dotsc, r_{(T-1)N+v} x_v,)$ can be viewed as a random code of length $mT$ bits. 
In other words, the proposed codes can be viewed as non-binary LDPC codes over $\GF(2^m)$ serially concatenated with $N$ random binary codes of length $mT$.
Intuitively, this explains why multiplicative repetition works better than simple repetition. 

The construction of the proposed codes  may remind some readers of Justesen codes \cite{1054893}. 
Note that the proposed construction chooses the multiplicative coefficients uniformly at random. 
Note also that since the minimum distance of $C_1$ is at most $O(\log(N))$ \cite{4641893}, the $C_T$ code has minimum distance is at most $O(T\log(N))$. 

Due to the repetition of symbols, the encoder  are inherently  rate-compatible.
\section{Decoding Scheme}
\label{223813_17Feb10}
The BP decoder for non-binary LDPC codes \cite{706440} exchanges probability vectors of length $2^m$, called {\it messages}, 
between variable nodes and check nodes, at each iteration round $\ell\ge 0$. 
The proposed codes $C_T$ for $T\ge 2$ also can be decoded by the BP decoding algorithm on the Tanner graphs of $C_T$. 
In this section, 
instead of the immediate use of the BP decoding on the Tanner graph of $C_T$, 
we propose a decoding algorithm which uses only the Tanner graph of $C_1$ for decoding $C_T$ for $T\ge 2$. 

\begin{figure}[t]
\begin{center}
  \includegraphics[scale=1.1]{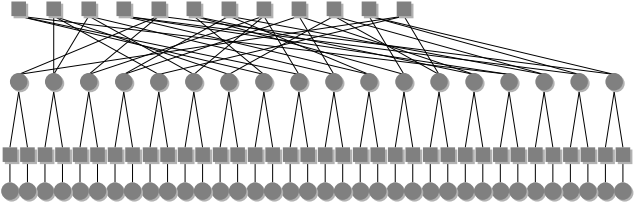}
 \caption{An example of $C_3$. A non-binary (2,3)-regular LDPC code over $\GF(2^m)$ concatenated with one inner multiplicative repetition codes of length 3. 
 The code length is 54 symbols or equivalently $54m$ bits.
 The rate is 1/9. }
 \label{175738_8Jan10}
\end{center} 
\end{figure}

The variable nodes of degree one in Fig.~\ref{174554_8Jan10} and Fig.~\ref{175738_8Jan10} represent multiplicative repetition symbols
of $C_2$ and $C_3$, respectively. 
If the BP decoding algorithm is immediately applied  to the proposed codes, 
all the variable nodes and check nodes, including the variable nodes of those multiplicative repetition symbols, are activated, i.e.~exchage the messages. 
However, the  messages reaching  the variable nodes of degree one do not change messages that are sent back from the nodes. 
Therefore,  the decoder does not need to pass the messages all the way to those variable nodes of degree 1 and 
their adjacent check nodes of degree 2.
Consequently, after the variable nodes of degree 1 pass the initial messages to the upper part of the graph, 
the decoder uses only the upper part of the graph, i.e. $C_1$. 

The computations of check nodes are more complex than those of variable nodes. 
As posed  in the Problem~1 in Section \ref{150913_18Feb10}, the number $M$ of the check nodes gets higher as $R$ decreases. 
In general, LDPC codes of information length $K$ and rate $R$ have $K(1-R)/R$ check nodes. In our setting, we have $K=N/3$ information symbols.
The number of check nodes in the proposed code $C_T$ for $T\ge 2$ is also given by $K(1-R)/R$. 
However, $(T-1)N$ check nodes of degree 2  adjacent to the $(T-1)N$ variable nodes of degree 1
do not need to participate in the BP decoding iterations. 
The only $2N/3$ active check nodes in the mother code $C_1$ participate in the BP decoding algorithm for decoding $C_T$ for $T\ge 2$. 
Note that the number of active check nodes $2N/3$ remains unchanged for any $T\ge 1$. 
This is highly preferable property for low-rate LDPC codes, which relieves the Problem~1. 
The Problem~2 is also relieved, since the maximum degree of check nodes in the mother code $C_1$ is as small as 3. 

The BP decoding involves mainly 4 parts, i.e.~the initialization, the check to variable computation, the variable to check computation, and the tentative decision parts. 
For $v=1,\dotsc,NT$, let $X_v$ be the random variables with realizations $x_v$. 
Let $Y_v$ be the random variables with realizations $y_v$ which is received value from the channel $\Pr(Y_v|X_v)$ 
and the probability of transmitted symbol $\Pr(X_v)$ is assumed to be uniform. 

We assume the decoder knows the channel transition probability
\begin{align}
 \Pr(X_v=x|Y_v=y_v), v=1,\dotsc,NT \label{223618_16Feb10}
\end{align} 
for $x\in \GF(2^m)$. 
When the transmissions take place over  the memoryless binary-input output-symmetric channels, 
we can rewrite \eqref{223618_16Feb10} as
\begin{align*}
&\Pr(X_v=x|Y_v=y_v)=\prod_{i=1}^m\Pr(X_{v,i}=x_i|Y_{v,i}=y_{v,i}),
\end{align*}
where
$(x_1,\dotsc,x_m)\in\GF(2)^m$ is the binary representation of $x\in\GF(2^m)$ and $X_{v,i}$ is the random variable of the transmitted bit, 
and the corresponding channel output $y_{v,i}$ and its random variable $Y_{v,i}$. 
\\\\
\subsection{Decoding Algorithm}
\initialization:\\
 For each variable node $v$ in $C_1$ for $v=1,\dotsc, N$, compute $p_{v}^{(0)}(x)$ as follows. 
 \begin{align}
\begin{split}
   p_{v}^{(0)}(x)&= \xi\Pr(X_{v}=x|Y_{v}=y_{v})\\
  &\prod_{t=1}^{T-1}{\Pr(X_{tN+v}=r_{tN+v}x|Y_{tN+v}=y_{tN+v})},
\end{split} \label{181252_11Jan10}
\end{align} 
 for  $x\in \GF(2^m)$, where $\xi$ is the normalization factor so that $\sum_{x\in\GF(2^m)}p^{(0)}_{v}(x)=1$.
 Each variable node $v=1,\dotsc, N$ in $C_1$ sends the initial message $p_{vc}^{(0)}=p_{v}^{(0)}\in\RR^{2^m}$ to 
 each adjacent check node $c$. 
 Set the iteration round as $\ell:=0$. 
\\\\\ctov:\\ 
For each check node $c=1,\dotsc,M$  in $C_1$,
let $\partial c$ be the set of the adjacent variable nodes of $c$. It holds that $\#\partial c=3$,   since the mother code $C_1$ is $(2,3)$-regular. 
Each $c$ has 3 incoming messages $p_{vc}^{(\ell)}$ for $v\in \partial c$ from the 3 adjacent variable nodes.
The check node $c$ sends the following message ${p}^{(\ell+1)}_{cv}\in\RR^{2^m}$ to each adjacent variable node $v\in \partial c$. 
\begin{align*}
&\tilde{p}^{(\ell)}_{vc}(x) = {p}^{(\ell)}_{vc}(h_{cv}^{-1} x) \text{ for $x\in\GF(2^m)$},\\
&\tilde{p}^{(\ell+1)}_{cv}= \otimes_{v'\in \partial c\setminus{\{v\}}}\tilde{p}^{(\ell)}_{v'c}, \\
&{p}^{(\ell+1)}_{cv}(x) = \tilde{p}^{(\ell+1)}_{cv}(h_{cv} x)\text{ for $x\in\GF(2^m)$}.
 \end{align*} 
where $p_1\otimes p_2\in\RR^{2^m}$ is a convolution of $p_1\in\RR^{2^m}$ and $p_2\in\RR^{2^m}$. To be precise, 
\begin{equation*}
  (p_1\otimes p_2)(x) = \sum_{\substack{y,z\in\GF(2^m)\\x=y+z}}{p_1(y)p_2(z)} \text{ for $x\in\GF(2^m)$}.
\end{equation*}
The convolution seems the most complex part of the decoding algorithm. 
Indeed, the convolutions are efficiently calculated via FFT and IFFT \cite{4155118}, \cite{RaU05/LTHC}.
Increment the iteration round as $\ell:=\ell+1$. 
\\
\\\vtoc: \\
Each variable node $v=1,\dotsc,N$ in $C_1$ has two adjacent check nodes since the mother code $C_1$ is $(2,3)$-regular. 
Let $c$ and $c'$ be the two adjacent check nodes of $v$. 
The message $p^{(\ell)}_{vc}\in\RR^{2^m}$ sent from $v$ to $c$ is given by
\begin{align*}
p^{(\ell)}_{vc}(x) = \xi p_v^{(0)}(x)p^{(\ell)}_{c'v}(x) \text{ for $x\in\GF(2^m)$},
 \end{align*} 
where $\xi$ is the normalization factor so that $\sum_{x\in\GF(2^m)}p^{(\ell)}_{vc}(x)=1$. \\
\\\tentative: \\
For each $v=1,\dotsc, N$, the tentatively estimated $v$-th transmitted symbol is given as 
\begin{align*}
\hat{x}_v^{(\ell)}=\mathop{\mathrm{argmax}}_{x\in \GF(2^m)}p_v^{(0)}(x)p^{(\ell)}_{cv}(x)p^{(\ell)}_{c'v}(x),
\end{align*}
where $c$ and $c'$ are the two adjacent check nodes of $v$. 
If $\underline{\hat{x}}^{(\ell)}:=(\hat{x}_1^{(\ell)},\dotsc,\hat{x}_N^{(\ell)})$ forms a codeword of $C_1$, in other words, 
  $\underline{\hat{x}}^{(\ell)}$ satisfies every  parity-check equations
\begin{align*}
 \sum_{v\in \partial c}h_{cv}\hat{x}_v^{(\ell)}=0\in \GF(2^m) 
\end{align*}
for all $c=1,\dotsc,M$, the decoder outputs $\underline{\hat{x}}^{(\ell)}$ as the estimated codeword. 
Otherwise repeat the latter 3 decoding steps. If the iteration round $\ell$ reaches a pre-determined 
number, the decoder outputs {\tt FAIL}.

The decoder is inherently rate-compatible. 
Indeed, for decoding the different  $C_T$ of rate $1/(3T)$ for $T=1,2,\dotsc$, 
the decoder only  needs the Tanner graph of the mother code $C_1$.

\section{Erasure Channel Analysis}
\label{224134_17Feb10}
In the binary case, we can predict the asymptotic decoding performance of LDPC codes transmitted over the general memoryless binary-input output-symmetric channels
in the large code length limit by {\it density evolution} \cite{richardson01design}. 
Density evolution also can be used to analyze non-binary LDPC codes \cite{1273653,4787626}. 
However,  for large field size, it becomes computationally intensive 
and tractable only for the BEC. 

Rathi and Urbanke developed the density evolution which enables the prediction of the decoding performance of the non-binary LDPC codes over the BEC in the limit of large code length. 
For a given code ensemble, density evolution gives the maximum channel erasure probability at which the decoding erasure probability, 
averaged over all the LDPC codes in the ensemble goes to zero. 
The maximum channel erasure probability given by the density evolution is referred to as {\itshape the threshold}. 

It is shown in \cite{RaU05/LTHC} that for the transmissions over the BEC with non-binary LDPC codes defined over $\GF(2^m)$, 
the decoding results depend on the binary representation, i.e.~the primitive element. 
In other words, two isomorphic fields do not, in general, yield the identical decoding results.
Rathi and Urbanke also observed that  the difference of the threshold is of the order of $10^{-4}$ for the different fields.
The density evolution \cite{RaU05/LTHC} is  developed for the non-binary LDPC code 
ensembles with parity-check matrices defined over the general linear group $\GL(\GF(2),m)$. 
In this section, we will use the density evolution to evaluate the thresholds of non-binary LDPC codes defined 
over $\GF(2^m)$. This is a fair approximation, since in \cite{RaU05/LTHC}, it is reported that the threshold for the code ensemble 
with parity-check matrices defined over $\GF(2^m)$ and $\GL(\GF(2),m)$  have  almost the same thresholds within the order of $10^{-4}$. 

When the transmission takes place over the BEC and all-zero codeword is assumed to be sent, the messages, described by probability vectors $(p(x))_{x\in\GF(2^m)}$ of length $2^m$ in general, 
can be reduced to linear subspaces \cite{RaU05/LTHC} of $\GF(2)^m$. 
To be precise, for each message in the BP decoding algorithm, a subset of $\GF(2)^m$
\begin{align*}
 \{\bm{x}\in \GF(2)^m\mid p(x)\neq 0\},
\end{align*}
forms a linear subspace of $\GF(2)^m$, where $\bm{x}$ is the binary representation of $x\in\GF(2^m)$.

Define $P^{(\ell)}=(P^{(\ell)}_0,\dotsc, P^{(\ell)}_m)$ and $Q^{(\ell)}=(Q^{(\ell)}_0,\dotsc,Q^{(\ell)}_m)$ as the probability vectors of length $m+1$ such that 
$P_i^{(\ell)}$ (resp.~$Q_i^{(\ell)}$) is 
the probability that a message sent from variable (resp.~ check) nodes has dimension $i$ at the $\ell$-th iteration round of the BP decoding algorithm. 
The density evolution gives us the update equations of $P^{(\ell)}$ and $Q^{(\ell)}$ for $\ell\ge 0$. 

  \begin{figure}[!t]
\begin{center}
    \includegraphics[width=\hsize]{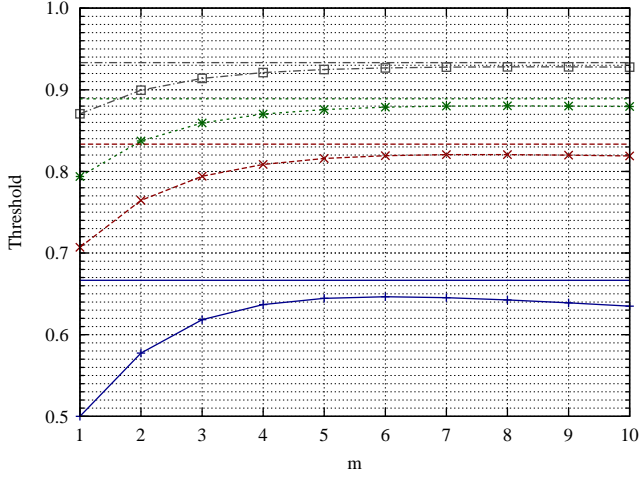}
   \caption{Thresholds $\epsilon^{\ast}$ of  $C_T$ over $\GF(2^m)$ for the BEC and  $T$=1,2,3 and 5 from below.
   The rate is $1/(3T)$. The straight lines show the Shannon limits  $1-1/(3T)$. }
  \label{140517_24Apr10}
\end{center}
\end{figure}

Rathi and Urbanke\cite{RaU05/LTHC} developed the density evolution for the BEC that tracks probability mass functions of the dimension of the linear subspaces. 
For $\ell\ge 0$, the density evolution tracks the probability vectors $P^{(\ell)}$ and $Q^{(\ell)}$
 which are referred to as {\itshape densities}. 
The initial messages in \eqref{181252_11Jan10} can be seen as the intersection of $T$ subspaces of the messages received as the channel outputs.
The density of the initial messages is given by $P^{(0)}$ as follows,
\begin{align*}
&P^{(0)}=\overbrace{E \boxdot\cdots\boxdot E}^{\text{$T$ times}},\\
& E:=(E_0,\dotsc, E_m),  \\
&E_i:=\binom{m}{i}\epsilon^{i}(1-\epsilon)^{m-i}, 
\end{align*}
where $\epsilon$ is the channel erasure probability of the BEC. 
The operator $\boxdot$ is defined as follows.
\begin{align*}
\left[ P \boxdot Q \right]_k &= \sum_{i=k}^m \sum_{j=k}^{k+m-i} C_{\boxdot }(m,k,i,j) P_i Q_j,\\
{C_\boxdot }(m,k,i,j) &:=2^{(i-k)(j-k)}\frac{\gb{i}{k}\gb {m-i} {j-k}  }{\gb m j }, 
\end{align*}
where $\displaystyle\gb{m}{k}=\prod_{l=0}^{k-1}\frac{2^m-2^l}{2^k-2^l}$
is a 2-Gaussian binomial. 

Since the mother code is (2,3)-regular, the update equations of density evolution is given by
\begin{align*}
& Q^{(\ell+1)}=P^{(\ell)}\boxtimes P^{(\ell)},\\
&  P^{(\ell+1)}=P^{(0)}\boxdot Q^{(\ell+1)}, 
\end{align*}
where the operator $\boxtimes$ is defined as follows.
\begin{align*}
\left[ P \boxtimes Q \right]_k &= \sum_{i=0}^k \sum_{j=k-i}^{k} C_{\boxtimes }(m,k,i,j) P_i Q_j,\\
 {C_\boxtimes }(m,k,i,j) &:=2^{(k-i)(k-j)} \frac{\gb {m-i} {m-k} \gb i {k-j} }{\gb m {m-j} }.
\end{align*}
Since the messages of dimension 0 corresponds to the successful decoding, the threshold is defined as follows. 
\begin{align*}
 &\epsilon^{\ast}:=\sup_{\epsilon\in[0,1]}\{\epsilon\in[0,1]\mid\lim_{\ell\to\infty}P^{(\ell)}_0=1\}.
\end{align*}
In the large code length limit, if $\epsilon<\epsilon^{\ast}$  the reliable transmissions are possible with the proposed $C_T$. 


 Fig.~\ref{140517_24Apr10} draws the thresholds of $C_T$ defined with parity-check matrices over $\mathrm{GL}_m(\GF(2))$
 for repetition parameter $T=1,\dotsc, 5$ and $m=1,\dotsc, 10$.
 The threshold  $\epsilon^{\ast}=1/\sqrt[T]{2}$ for the binary case $m=1$ is decided by the stability condition  \cite{mct}.
 It can be seen that the thresholds are not monotonic with respect to $m$.
 For repetition parameter $T=1$, i.e., the mother code has the maximal threshold at $m=6$.
 For $T>2$, the maximal threshold is attained around at $m=8$.

\begin{figure}[!t]
\begin{center}
 \includegraphics[width=\hsize]{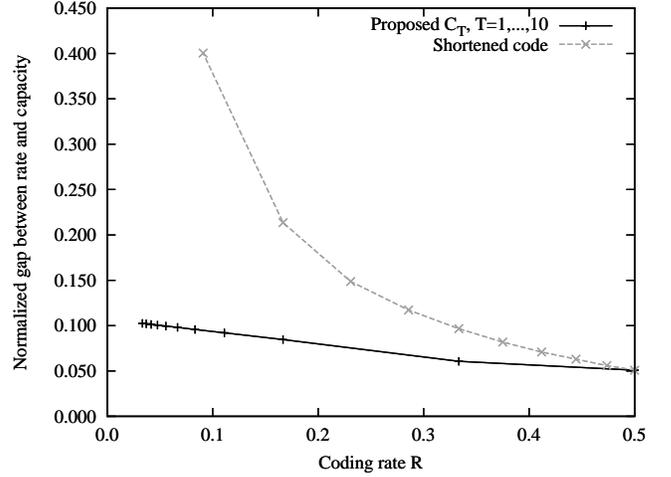} 
 \caption{
 The asymptotic decoding performance over the BEC of the proposed codes and the best known low-rate codes. 
 One curve corresponds to  the proposed code $C_T$ over $\GF(2^6)$ of coding rates $R=1/(3T)$ with repetition parameter $T=1,\dotsc,10$. 
 The punctured $C_1$ of rate 1/2 is also plotted. 
 The other curve corresponds to the bit-wise shortened non-binary LDPC code over $\GF(2^6)$ proposed by Klinc et al.~\cite[Fig.~1]{4797675}. 
 The vertical axis indicates $(1-\epsilon^{\ast}-R)/R$ which is the normalized gap between the capacity $1-\epsilon^{\ast}$ and the rate $R$, where $\epsilon^{\ast}$ is the threshold.
} 
 \label{134623_18Feb10}
\end{center}
\end{figure}

Fig.~\ref{134623_18Feb10} compares the proposed codes and the best existing low-rate LDPC codes respect to the thresholds for the BEC. 
It can be seen that the proposed codes have better thresholds especially for lower rates. 

  \begin{figure}
\begin{center}
     \includegraphics[width=\hsize]{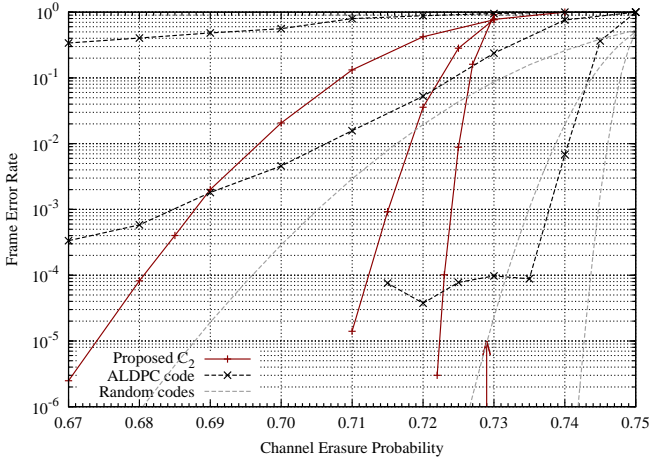}
 \caption{
The solid curve shows the decoding performance of the proposed code $C_2$ whose mother code is a $(2,4)$-regular non-binary LDPC code defined over $\GF(2^8)$. 
The coding rate is 1/4. 
The transmission takes place over the BEC. The arrow indicates the threshold 0.72898 of $C_2$. 
The code length is of length 1024, 8192 and 65536. 
For comparison,  the decoding performance of accumulated LDPC (ALDPC) codes \cite[Fig.~15]{4215147} is shown. 
It is known that the ALDPC codes achieve the capacity of the BEC in the limit of large code length and exhibit good decoding performance with  finite code length.
The frame error rate  of corresponding random codes of rate 1/4 under maximum-likelihood decoding are calculated by \cite[Eq.~(3.2)]{1003839}.
It can be seen that the proposed codes exhibit better decoding performance than the ALDPC codes with code length up to 8192. 
The error floors of the proposed codes can not be observed down to FER $10^{-5}$ while the ALDPC codes have high error floors even with code length as long as 65536 bits.
}
 \label{224151_25Nov10}
\end{center}   
\end{figure}

  \begin{figure}
\begin{center}
    \includegraphics[width=\hsize]{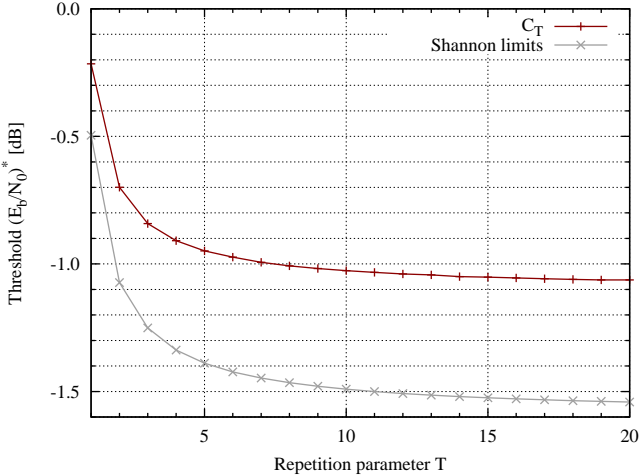}
   \caption{
The thresholds for the AWGN channels of the proposed codes $C_T$ of rate $1/(3T)$ for $T=1,\dotsc,20$. The codes are defined over $\GF(2^8).$
}
   \label{224517_25Nov10}
\end{center}
\end{figure}     

  \begin{figure}
\begin{center}
     \includegraphics[width=\hsize]{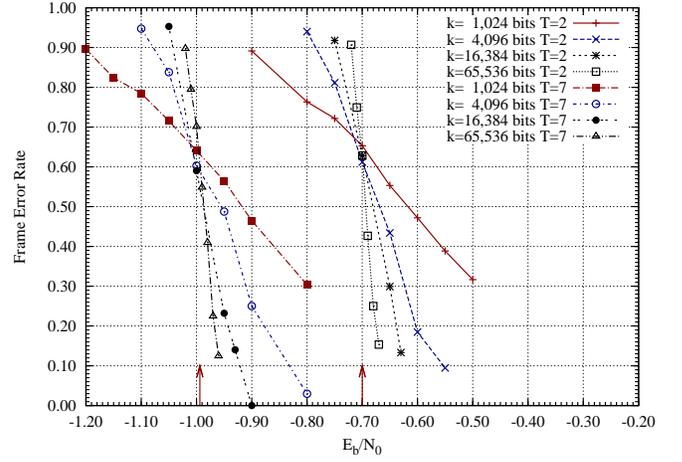}
     \caption{
Frame error rate versus parameter for the proposed codes $C_2$ and $C_7$ over $\GF(2^8)$ transmitting over the AWGN channel.
The rates of $C_2$ and $C_7$ are 1/6 and 1/21, respectively. 
The information length are set to 1024, 4096, 16834, and 65536. 
The arrows indicate the corresponding threshold values. 
Observe how the curves move closer to these threshold values for increasing codeword lengths.
}
     \label{132144_26Nov10}
\end{center}   
\end{figure}

  \begin{figure}
\begin{center}
      \includegraphics[width=\hsize]{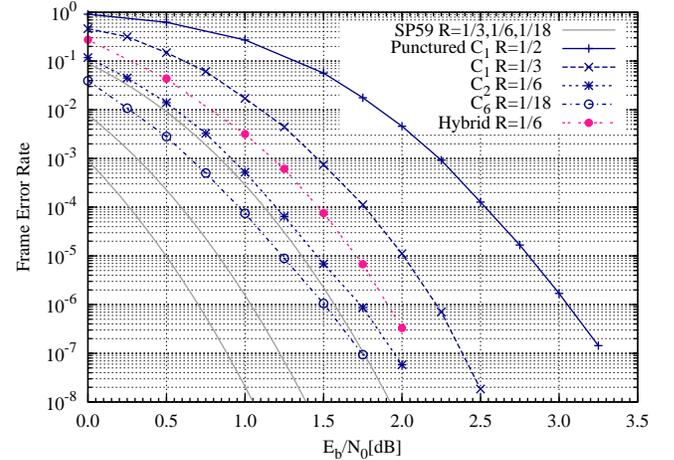}
     \caption{The frame error rate of the proposed codes $C_T$ for $T=1,2,6$ and hybrid non-binary LDPC codes. 
 It also shows the performance of rate half punctured mother code $C_1$. All these codes have 192 information bits.
The curves labeled SP59 are the corresponding Shannon's 1959 sphere-packing bound \cite{shannon59,1362893,4494709} for rate 1/3, 1/6 and 1/18.
}
     \label{065003_12Jan10}
\end{center}     
   \end{figure}
  \begin{figure}
\begin{center}
     \includegraphics[width=\hsize]{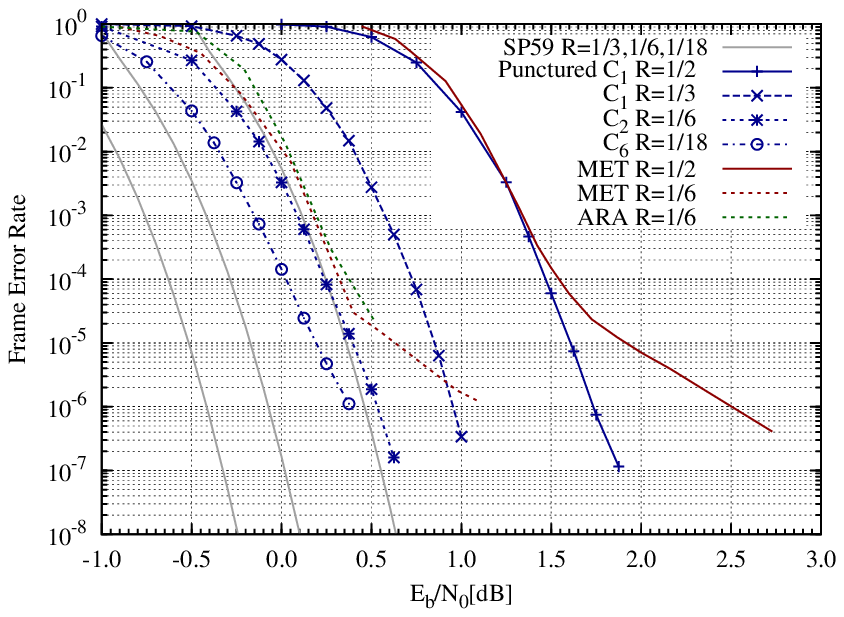}
     \caption{The frame error rate of the proposed codes $C_T$ for $T=1,2,6$, multi-edge type (MET) LDPC code of rate 1/2 \cite{met} and 1/6,  and accumulate repeat accumulate (ARA) code \cite{1523619}  of rate 1/6.  
All these codes have 1024 information bits except that the MET LDPC code of rate 1/2 has 1280 information bits. It also shows the performance of rate half punctured mother code $C_1$.
The curves labeled SP59 are the corresponding Shannon's 1959 sphere-packing bounds for rate 1/3, 1/6 and 1/18.
}
     \label{093937_12Jan10}
\end{center}   
\end{figure}
\section{Numerical Results}
\label{224143_17Feb10}
In this section, we give some numerical results of the proposed codes. 
Fig.~\ref{224151_25Nov10} shows the decoding performance of the proposed code $C_2$ whose mother code is a $(2,4)$-regular non-binary LDPC code defined over $\GF(2^8)$.  
The transmission takes place over the BEC. 
The compared accumulated LDPC (ALDPC) codes are designed to achieve the capacity in the limit of large code length. 
It can be seen that the proposed codes exhibit better decoding performance than the ALDPC codes with code length up to 8192. 
The error floors of the proposed codes can not be observed down to frame error rate $10^{-5}$ while the ALDPC codes have high error floors even with code length as long as 65536 bits.

Fig.~\ref{224517_25Nov10} shows the  thresholds of the proposed codes $C_T$ of rate $1/(3T)$ for $T=1,\dotsc,20$. The codes are defined over $\GF(2^8).$ 
The threshold values are calculated by the Monte Carlo simulation method. The method was originally suggested in \cite[p.~22]{davey_phd_thesis} and an efficient calculation was developed  in \cite[Section VII]{5571892}.
It can be observed that  the proposed codes leave a gap to the ultimate Shannon limit $\mathrm{E_b/N_0}=\log_{10}(\ln(2))\approx -1.59$ [dB] \cite{haykin01} even  in the limit of large repetition parameter $T$. 
Fig.~\ref{132144_26Nov10} depicts the simulation results for the AWGN channels of $C_2$ and $C_7$ of very long code length. 
We observe the convergence 
to a sharp threshold effect at the predicted threshold values as the information length $k$ increases.

We demonstrate the  decoding performance of the short and moderate-length proposed codes $C_T$ for $T=1,2,3,4,6$ over the binary-input AWGN channels. 
The mother code $C_1$ is constructed by the optimization method in \cite{4641893}. 
The coefficients $r_{N+1},\dotsc, r_{TN}$ are chosen uniformly at random from $\GF(2^m)\setminus\{0,1\}$, 
where $1\in\GF(2^m)$ is the multiplicative identity. 
We fix $m=8$ for its good performance and the computer-friendly representation of byte. 

   Fig.~\ref{065003_12Jan10} shows the decoding performance of $C_T$ for $T=1,2,3,6$ of rates ${1}/{(3T)}$. It also shows a hybrid non-binary LDPC code \cite{4658702} of rate 1/6 and punctured $C_1$ of rate $1/2$. 
   All these codes have 192 information bits. 
   The proposed code $C_2$ outperforms the hybrid non-binary LDPC code which is the best code  so far for that rate and  code length. 
   The code $C_3$ of rate 1/9 has about 0.5 [dB] coding gain from $C_2$ of rate 1/6. 
As we show on these curves, the proposed construction, although simple, allows to design codes 
with very low rates without large loss of gap to the Shannon limits. 

   The same property can be seen for the proposed codes with larger information bits. 
   Fig.~\ref{093937_12Jan10} shows the decoding performance of $C_T$ for $T=1,2,3,6$, the binary multi-edge type  LDPC code of rate 1/2 and 1/6, and the binary ARA code \cite{1523619} of rate 1/6. 
   All these codes have 1024 information bits
except that the MET LDPC code of rate 1/2 has 1280 information bits. 
  It also shows the performance of a punctured mother code $C_1$ of rate 1/2.
   Among the  codes of rate 1/6, the proposed code  $C_2$ has the best performance both at water-fall and error-floor regions. 

As posed in Problem 3, conventional low-rate codes required large code length to exploit the potential performance. 
It can be seen that the proposed codes exhibit better decoding performance both at small and moderate code length. 
\section{Conclusions}
We propose non-binary LDPC codes concatenated with inner multiplicative repetition codes. 
The performance of the proposed codes exceeds the hybrid non-binary codes, multi-edge type LDPC codes and ARA codes both at the water-fall and error-floor regions. 
The encoder and decoder are inherently rate-compatible, and especially the decoder complexity is almost the same as the mother code. 
\section*{Acknowledgments}
The authors are grateful to T.~Richardson for providing the data of the multi-edge type LDPC code of rate 1/6 in Fig.~\ref{093937_12Jan10}. 
The authors would like to thank anonymous reviewers of ISIT2010 and Transactions on Information Theory, and I.~Sason for their suggestions and comments. 
K.~K.~  wishes to thank T.~Uyematsu  for valuable comments. 
\bibliographystyle{IEEEtran}
\bibliography{IEEEabrv,kenta_bib}

\end{document}